\documentclass[12pt]{article}
\newcommand{\be}{\begin{equation}}
\newcommand{\ee}{\end{equation}}

\newcommand{\bi}{\begin{itemize}}
\newcommand{\ei}{\end{itemize}}
\newcommand{\bea}{\begin{eqnarray}}
\newcommand{\eea}{\end{eqnarray}}
\newcommand{\ba}{\begin{array}}
\newcommand{\ea}{\end{array}}
\newcommand{\IR}{\mathbb{R}}
\newcommand{\nn}{\nonumber}
\usepackage[left=2.50cm, right=2.50cm, top=2.50cm, bottom=2.50cm]{geometry}
\usepackage[utf8]{inputenc}
\usepackage{cancel}
\usepackage{placeins}
\usepackage{bm}
\usepackage[english]{babel}
\usepackage{physics}
\usepackage{hyperref}
\usepackage[sort]{cite}
\usepackage{amsthm}
\usepackage{mathrsfs}
\usepackage{mathtools}
\usepackage{graphicx}
\usepackage{changepage}
\usepackage{amssymb,amsmath}
\usepackage{verbatim}
\usepackage{empheq}
\usepackage{tikz}
\usepackage{pgfplots}
\pgfplotsset{compat=1.12}
\usepackage{float}
\usepackage{multirow}
\usepackage{multicol}
\usepackage{amsmath}
\usepackage{centernot}
\usepackage[hang,flushmargin]{footmisc} 
\usepackage[normalem]{ulem}
\usetikzlibrary{tikzmark}
\numberwithin{equation}{section}
\hypersetup{hidelinks}

\newlength{\bibitemsep}
\newlength{\bibparskip}
\let\oldthebibliography\thebibliography
\renewcommand\thebibliography[1]{%
  \oldthebibliography{#1}%
  \setlength{\parskip}{\bibitemsep}%
  \setlength{\itemsep}{\bibparskip}%
}
\begin{document}
\thispagestyle{empty}

\begin{minipage}{0.92\textwidth}
{\LARGE\bfseries
Covariant Fracton Electrodynamics in Six Dimensions\par}

\vspace{0.35em}

\vspace{1.0em}
{\large\bfseries Nicola Maggiore$^{a,b}$\textsuperscript{,}\footnotemark\par}

\vspace{0.6em}
{\small
$^{a}$ Dipartimento di Fisica, Universit\`a di Genova, Via Dodecaneso 33, I-16146 Genova, Italy\par
$^{b}$ Istituto Nazionale di Fisica Nucleare -- Sezione di Genova, Via Dodecaneso 33, I-16146 Genova, Italy\par
}
\end{minipage}

\footnotetext{\texttt{\href{mailto:nicola.maggiore@ge.infn.it}{nicola.maggiore@ge.infn.it}}}

\vspace{0.6em}

\begin{center}
\begin{minipage}{0.92\textwidth}
\noindent{\bfseries Abstract.}\;
We formulate a covariant version of Maxwell-like fracton electrodynamics in six dimensions using a symmetric tensor gauge field with scalar gauge symmetry $\delta A_{\mu\nu}=\partial_\mu\partial_\nu\Lambda$. This provides a relativistic setting in which the characteristic fractonic restriction on mobility follows directly from gauge invariance and the allowed coupling to matter. We construct the stress--energy tensor and show that its trace has a universal dimension-dependent structure that becomes a total derivative in $d=6$. In the presence of sources, the theory enforces conservation of charge and dipole moment, capturing the immobility of isolated charges and the mobility of dipolar bound states. This structure can also be viewed as a higher-moment form of generalized global symmetry.
\end{minipage}
\end{center}

\vspace*{0pt plus 1fill}
\vspace{\baselineskip}

\begin{sloppypar}
\end{sloppypar}

\newpage

\section{Introduction}

We study scalar-charge fracton electrodynamics in a manifestly covariant setting
based on a symmetric rank-2 gauge field \(A_{\mu\nu}\) and the scalar gauge symmetry
\(\delta A_{\mu\nu}=\partial_\mu\partial_\nu\Lambda\).
One motivation is to place the theory in a relativistic setting   in which  the
stress--energy tensor, the coupling to external sources, and the origin of the
mobility constraints can be analyzed   within a single framework , without importing
nonrelativistic restrictions by hand.
A key outcome is a universal, dimension-dependent trace structure for the
stress--energy tensor: in \(d=6\) the trace reduces to a total derivative,
signalling classical scale invariance in flat space.
This does not automatically imply the existence of a local, manifestly
gauge-invariant traceless representative at the level of the stress--energy
tensor: within the usual class of local gauge-invariant trace improvements
this is obstructed, although a traceless representative can be found upon
imposing the equations of motion.
The interest of the six-dimensional setting is not that it should be taken as a
direct model of observed spacetime physics. Rather, it identifies the dimension
in which the present tensor gauge structure acquires its simplest local
Wilsonian realization: the basic two-derivative action is marginal, and the
scaling assignment of the gauge parameter matches the standard field-theoretic
expectation for a gauge symmetry.

 From this viewpoint, covariance is useful not because fracton phases are intrinsically relativistic, but because it isolates the symmetry principle responsible for the restricted-mobility sector. In particular, it cleanly separates what follows directly from the local gauge symmetry itself---admissible source couplings, conserved multipole moments, and the associated mobility rules---from what instead depends on specific nonrelativistic realizations. This is the sense in which the present construction should be read as a complementary field-theoretic framework, rather than as a replacement for the standard condensed-matter descriptions.

 The gauge symmetry also constrains the admissible sources \(J^{\mu\nu}\):
consistency requires \(\partial_\mu\partial_\nu J^{\mu\nu}=0\), implying
conservation of charge and dipole moment and providing a minimal entry point
to higher-moment generalized global symmetries.
A characteristic feature of fracton phases is the presence of excitations
whose mobility is constrained by multipole conservation laws, most notably
the coexistence of conserved charge and conserved dipole moment.
In many effective descriptions these constraints are imposed directly in a
nonrelativistic setting~\cite{Gromov:2020yoc}, or encoded in
higher-rank gauge structures where the gauge potential is a spatial symmetric
tensor~\cite{Pretko:2016kxt,Pretko:2016lgv}.
A covariant formulation is attractive because it makes  both  the symmetry origin
and the domain of validity of the fractonic constraints manifest: it specifies
which source/background couplings are compatible with a consistent fractonic
sector, rather than appealing to nonrelativistic intuition only {\it a posteriori}.
It also provides   a compact organization of the  allowed deformations.

Early examples of constrained dynamics and ``fractonic'' behavior appeared both
in lattice models of topological overprotection and in stabilizer-code
constructions, including Chamon's quantum glassiness
model~\cite{Chamon:2004lew} and later fractal codes such as Haah's cubic
code~\cite{Haah:2011drr} and its descendants~\cite{Bravyi:2011faf,Williamson:2016jiq}.
Related microscopic realizations and duality constructions include two-spin
Hamiltonians and generalized lattice gauge
theories~\cite{Slagle:2017jgw,Vijay:2016phm}.

A particularly influential continuum perspective is provided by rank-two \(U(1)\)
gauge theories and their generalized electromagnetism, which capture
subdimensional particle mobility and multipole conservation directly at the
level of gauge
symmetry~\cite{Pretko:2016kxt,Pretko:2016lgv,Bulmash:2019taq}.
Related continuum constructions and symmetry-based formulations were developed
in~\cite{Radzihovsky:2019jdo,Seiberg:2020bhn,Gorantla:2020xap} and
further elaborated
in~\cite{Gorantla:2022eem,Gorantla:2022ssr}.
These structures organize fracton phases and their field-theoretic deformations,
including Higgs/partial confinement mechanisms and fractonic
criticality~\cite{Ma:2018nhd,Prem:2018jsn,You:2019cvs}.

Another complementary viewpoint emphasizes subsystem and fractal symmetries
and their gauging, leading to foliated fracton order and corresponding continuum
foliated field theory
descriptions~\cite{Shirley:2018vtc,Slagle:2018swq,Slagle:2020ugk,Gorantla:2025fzz}.
Algebraic approaches based on multipole symmetry and the associated operator
algebras have also been developed~\cite{Gromov:2018nbv}.
For broader context and additional perspectives on fracton phases, see the
reviews and
primers~\cite{Pretko:2020cko,Nandkishore:2018sel}.
Additional discussions of lattice models, numerics, and phenomenology,
including circuit and disorder phenomena, can be found
in~\cite{Pai:2018gyd,McGreevy:2022oyu}.
 For earlier general discussions of gauge procedures involving rank-two and higher-rank gauge fields beyond the specific fracton setting, see for example Ref.~\mbox{\cite{Singleton:2004je}}\hskip0pt.

 To our knowledge, a covariant scalar-charge fracton gauge theory in six spacetime dimensions has not been studied previously. The six-dimensional setting is nevertheless of broader theoretical interest. In relativistic field theory, six dimensions are well known to provide a distinguished arena for scale and conformal structures, as illustrated for example by six-dimensional superconformal theories~\mbox{\cite{Seiberg:1996qx,Heckman:2013pva,Heckman:2018jxk,Cordova:2015fha}}\hskip0pt. We do not suggest that the present symmetric-tensor fracton theory belongs to those classes. Rather, these examples indicate more generally that $d=6$ often plays a structurally special role when locality, gauge symmetry, and scaling are tightly constrained. In the present case, this role is tied to the fact that, under the standard assignment $[\Lambda]=0$, the local two-derivative Maxwell-like action is marginal precisely in six dimensions. This six-dimensional setting may therefore be viewed as a natural Wilsonian reference point for the present tensor gauge structure.

 Covariant treatments closely related in spirit have been developed in other dimensions,
in particular in $d=4$ for the Maxwell-like theory and its relation to linearized
gravity~\cite{Bertolini:2022ijb,Blasi:2022mbl,Bertolini:2023juh}. The present work focuses instead on
$d=6$, which is the unique dimension in which power counting makes a local two-derivative
kinetic term marginal while allowing the scalar gauge parameter to remain dimensionless.
This makes $d=6$ the natural dimension in which the covariant tensor gauge theory can be
examined at its most economical local fixed point, rather than as an effective theory already
dressed by a dimensionful coupling. For this reason, the six-dimensional setting will be our
preferred reference point throughout.
 Throughout, whenever we say that $d=6$ is singled out, this statement is understood under the standard assignment $[\Lambda]=0$. In this qualified sense, the role of six spacetime dimensions is Wilsonian and structural: it identifies the simplest local covariant fixed point of the tensor gauge symmetry, rather than a phenomenological claim about physical spacetime.

We therefore consider a covariant tensor gauge theory based on a symmetric
rank-2 field \(A_{\mu\nu}=A_{\nu\mu}\) with gauge symmetry
\be
\delta_{\rm fract}A_{\mu\nu}=\partial_\mu\partial_\nu\Lambda,
\label{intro_fract}
\ee
where \(\Lambda\) is a scalar local gauge parameter. The transformation \eqref{intro_fract} can be obtained by the ordinary infinitesimal diffeomorphism
\be
\delta_\xi A_{\mu\nu}=\partial_\mu\xi_\nu + \partial_\nu\xi_\mu
\label{diff}
\ee
by requiring the vector parameter \(\xi_\mu\) to be the gradient of a scalar
\be
\xi_\mu=\frac{1}{2}\partial_\mu\Lambda.
\label{xidiff}
\ee
For this reason the transformation \eqref{intro_fract} is sometimes referred to as ``longitudinal diffeomorphism''.
We adopt the standard field-theory convention of a dimensionless gauge parameter,
\be
[\Lambda]=0,
\label{Lambda0}
\ee
which fixes
\be
[A_{\mu\nu}]=2.
\label{Adim2}
\ee
With this assignment, a local two-derivative kinetic action of schematic form
\(\int d^dx\,\partial A\,\partial A\) has mass dimension \(d-6\), and is therefore
marginal only at \(d=6\). In this sense \(d=6\) occupies for the present tensor
gauge symmetry the same structural position that \(d=4\) occupies for ordinary
Maxwell theory.

We develop the corresponding theory in \(d=6\) and make explicit several structures
that are particularly transparent in this setting.
First, we classify the leading gauge-invariant deformations and reduce them to
an independent set using integrations by parts, Bianchi identities, and
equation-of-motion equivalences.
Second, we construct the symmetric stress--energy tensor and analyze its trace:
the universal \((d-6)\) structure implies that at \(d=6\) the generalized rank-3
field-strength term drops out and the trace reduces to a total derivative.
We then ask how far trace improvement can be pushed at the level of a local,
manifestly gauge-invariant stress--energy tensor, and we isolate the corresponding
obstruction, while showing that a traceless representative nevertheless exists
upon imposing the equations of motion, in the spirit of the standard analysis of
Callan--Coleman--Jackiw (CCJ)~\cite{Callan:1970ze}.
Third, we present a \(5+1\) decomposition that yields a compact generalized
Maxwell-like system for gauge-invariant electric and magnetic variables, both in
vacuum and in the presence of sources, reproducing the hallmark mobility
constraints in a covariant framework.

Our results in \(d=6\) complement existing treatments in other dimensions, where
the same gauge symmetry is naturally viewed as defining an effective field theory
with a dimensionful coupling.
For background on scalar-charge fracton theories and their continuum realizations
we refer to the gauge-principle
construction~\cite{Pretko:2018jbi}, the elasticity
duality~\cite{Pretko:2017kvd}, and continuum QFT approaches
highlighting higher-moment global
structures~\cite{Gromov:2017vir,Seiberg:2020bhn}.
For a broader perspective on generalized global symmetries we point to
Refs.~\cite{Gaiotto:2014kfa,Lake:2018dqm,McGreevy:2022oyu}.

In this work, we emphasize the following points within a unified covariant treatment:
\begin{itemize}
\item[(i)] a six-dimensional naturalness perspective with a dimensionless scalar
gauge parameter, which singles out \(d=6\) as the unique dimension in which power
counting makes a local two-derivative kinetic term marginal while keeping the
gauge parameter dimensionless;
\item[(ii)] a systematic construction of the stress--energy tensor, including the
universal \((d-6)\) trace structure and a gauge-invariant obstruction to a local,
manifestly gauge-invariant traceless improvement;
\item[(iii)] an explicit \(5+1\) constraint analysis and degree-of-freedom count for
the covariant theory, together with a compact Maxwell-like \(E/B\) system and its
coupling to sources.
\end{itemize}

The paper is organized as follows.
In Sect.~\ref{setup} we introduce the generalized higher-rank field
strength \(F_{\mu\nu\rho}\) and construct the most general invariant action
compatible with the power counting.
The Wilsonian operator classification and the basis relations among
gauge-invariant deformations are discussed in Sect.~\ref{wilson}.
Sect.~\ref{5plus1} develops the \(5+1\) decomposition in terms of
gauge-invariant electric and magnetic variables, derives the generalized
Maxwell system, and shows how coupling to sources enforces the mobility
constraints via conservation of charge and dipole moment.
The construction of the stress--energy tensor, together with the analysis
of its trace and the question of possible improvements, is carried out in
Sect.~\ref{stress}.
Sect.~\ref{conc} collects our conclusions and outlook.
The appendices contain complementary technical details.

\section{Covariant theory: field strength and minimal action}
\label{setup}

A strictly invariant field-strength-like tensor is obtained from the symmetric three-term combination,
\be
F_{\mu\nu\rho}\equiv F_{\nu\mu\rho}
= \partial_\mu A_{\nu\rho}+\partial_\nu A_{\mu\rho}-2\partial_\rho A_{\mu\nu}.
\label{defF}
\ee
It is symmetric in $(\mu,\nu)$ and obeys the cyclic identity
\be
F_{\mu\nu\rho}+F_{\nu\rho\mu}+F_{\rho\mu\nu}=0,
\label{cyclic}
\ee
as a direct consequence of the symmetry of the tensor gauge field $A_{\mu\nu}$. A Bianchi-type identity also holds identically:
\be
\partial_\sigma\bigl(F_{\mu\rho\nu}-F_{\nu\rho\mu}\bigr)+\partial_\mu\bigl(F_{\nu\rho\sigma}-F_{\sigma\rho\nu}\bigr)+\partial_\nu\bigl(F_{\sigma\rho\mu}-F_{\mu\rho\sigma}\bigr)=0.
\label{bianchi}
\ee
Throughout, we use the three-term gauge-invariant rank-3 tensor \eqref{defF}. Useful identities in this convention are collected in App.~\ref{dictionary}.

We consider the Maxwell-like action
\be
S_{\rm fract}=-\frac{1}{12}\int d^6x\,F_{\mu\nu\rho}F^{\mu\nu\rho},
\label{Sfract}
\ee
in $d=6$ (the overall normalization can be absorbed by a field rescaling).
Under the strict requirement $\dim(\mathcal L)\leq 6$ in six dimensions and locality, gauge invariance restricts the leading action to contain exactly two derivatives: terms with fewer derivatives are not gauge invariant, while higher-derivative terms have $\dim(\mathcal L)>6$ and are therefore excluded by the power counting.
Accordingly, the most general invariant functional respecting power counting is necessarily spanned by two independent structures. Besides $F_{\mu\nu\rho}F^{\mu\nu\rho}$ one can form the gauge-invariant trace vector
\be
K_\mu\equiv \partial^\alpha A_{\alpha\mu}-\partial_\mu A = -F_{\alpha\mu}{}^{\alpha},\qquad A\equiv A^{\alpha}{}_{\alpha},
\label{Kdef}
\ee
and its square $K_\mu K^\mu$.
A second contraction $F_{\mu\nu\rho}F^{\mu\rho\nu}$ is not independent for a symmetric $A_{\mu\nu}$: one finds the exact identity
\be
F_{\mu\nu\rho}F^{\mu\rho\nu}=-\tfrac12\,F_{\mu\nu\rho}F^{\mu\nu\rho},
\label{I2halfI1}
\ee
which follows immediately from the symmetry $A_{\mu\nu}=A_{\nu\mu}$ and from the cyclic identity \eqref{cyclic}.

Therefore the most general action compatible with the power counting reads
\be
S_0=-\frac{1}{12}\int d^6x\,\Big(F_{\mu\nu\rho}F^{\mu\nu\rho}+3\kappa\,K_\mu K^\mu\Big),
\label{Sgen}
\ee
with a single dimensionless parameter $\kappa$.
A particularly relevant choice is the linearized Einstein--Hilbert action, which in our notation reads
\be
\mathcal L_{\rm LG}=-\frac{1}{12}\,F_{\mu\nu\rho}F^{\mu\nu\rho}+\frac{1}{2}\,K_\mu K^\mu,
\label{FPinvariants}
\ee
corresponding to $\kappa=-2$ in \eqref{Sgen}.   At the level of the quadratic Lagrangian, this identifies the unique value of $\kappa$ for which the action matches the Einstein--Hilbert kinetic term. By itself, however, this coincidence does not enlarge the scalar gauge symmetry \eqref{intro_fract} to full linearized diffeomorphism invariance \eqref{diff}; the latter must be imposed separately .
We will use the simplest Maxwell-like representative \eqref{Sfract} as a convenient reference action, and treat the remaining quadratic structure encoded in $K_\mu K^\mu$ as an allowed two-derivative deformation.

We work in Minkowski signature $\eta_{\mu\nu}=\mathrm{diag}(-,+,+,+,+,+)$. With this choice and the overall sign in \eqref{Sfract}, the branch Hamiltonian discussed in Sec.~\ref{5plus1} is positive.
We explicitly verify the sign by computing $T_{00}$ in Sec.~\ref{stress} and again in App.~\ref{hamiltonian}.
 Our scope is correspondingly limited: all statements below refer to the Maxwell-like theory defined entirely by the gauge-invariant field strength $F_{\mu\nu\rho}$ , corresponding to $\kappa=0$ in the general family \eqref{Sgen}. In the general family, the instability identified in~\mbox{\cite{Afxonidis:2024tph}}\hskip0ptarises for kinetic structures involving the $K_\mu K^\mu$ term with values of $\kappa$ that generate ghost-like modes. At $\kappa=0$, the branch Hamiltonian density is positive semi-definite as established directly in Sect.~\ref{stress} and App.~\ref{hamiltonian} . We do not assume that    this positivity extends to the broader family; for related stability analyses see also~\mbox{\cite{Alvarez:2006uu}}\hskip0pt .
 Varying \eqref{Sfract} with respect to $A_{\alpha\beta}$ yields the Maxwell-type field equations
\be
\partial_\mu F^{\alpha\beta\mu}=0.
\label{eom}
\ee
The Bianchi identity \eqref{bianchi} supplies the complementary kinematic relations, in direct analogy with electromagnetism.  Here and in the following, ``Maxwell-like'' refers only to the two-derivative structure built from a gauge-invariant field strength together with its Bianchi identity; it does not imply equality of gauge content or propagating spectrum with ordinary electrodynamics.

With the mass dimension assignment \eqref{Adim2}, the kinetic term \eqref{Sfract} is marginal.
Equivalently, $[F_{\mu\nu\rho}]=3$ and $[F_{\mu\nu\rho}F^{\mu\nu\rho}]=6$.
 This statement should be understood under the standard field-theoretic assignment of a dimensionless scalar gauge parameter, namely $[\Lambda]=0$.
 In this sense $d=6$ is a natural Wilsonian reference point: with $[\Lambda]=0$ and $[A_{\mu\nu}]=2$, the two-derivative kinetic term is marginal, so higher-derivative local gauge-invariant deformations are naturally ordered by their mass dimension relative to the two-derivative theory.
Moreover, within locality and power counting, there are no nontrivial local gauge-invariant interaction terms at the leading two-derivative order (cubic, quartic, \emph{etc.}); the leading theory is therefore Gaussian, much as in linearized gravity.
Sect.~\ref{wilson} makes this organization concrete by classifying local invariants and isolating a minimal independent set, using integrations by parts, Bianchi identities, and equations of motion.

It is instructive to relate the Maxwell-like scalar-gauge theory developed here to the
better-known linearized Einstein theory in six dimensions (see e.g.\ Ref.~\cite{VanNieuwenhuizen:1973fi} for a general discussion of linearized Einstein gravity in arbitrary dimension).

As already noted, the gauge transformation \eqref{intro_fract}
can be viewed as the longitudinal subset of linearized diffeomorphisms \eqref{diff} .  For the specific value $\kappa=-2$ in the general quadratic action \eqref{Sgen}, the   quadratic Lagrangian coincides with the linearized Einstein--Hilbert kinetic term. This should not, however, be read as full equivalence of the two theories: the present model still retains only the scalar gauge symmetry \eqref{intro_fract}, whereas ordinary linearized gravity is characterized by the full vector gauge invariance \eqref{diff}. Only after enlarging the gauge symmetry from \eqref{intro_fract} to \eqref{diff} does one recover  standard six-dimensional linearized   gravity .

The six-dimensional setting emphasizes the conceptual distinction between the two theories.
Ordinary linearized gravity possesses no propagating scalar trace mode: this follows from the full linearized diffeomorphism invariance together with the associated constraints.
By contrast, in the present Maxwell-like theory the trace
$A^{\mu}{}_{\mu}$ is physical: it survives all gauge redundancies and contributes a scalar
polarization absent in linearized gravity.  
This difference is directly reflected in the
5+1 decomposition of Sec.~4, where the transverse tensor sector matches the nine polarizations
of a six-dimensional graviton, while the additional scalar mode is characteristic of the
fractonic gauge structure.

This comparison clarifies that the fracton-like Maxwell theory should not be interpreted as a
simple linearized gravitational system, but as a distinct covariant gauge structure whose
physical content and mobility constraints differ qualitatively from those of spin-2 dynamics.

As we will see shortly, invariance of the theory under \eqref{intro_fract} implies the covariant constraint for a symmetric source,
$\partial_\mu\partial_\nu J^{\mu\nu}=0$,
which in turn yields conservation of the total charge and of the total dipole moment in $\IR^5$, as reviewed in Sect.~\ref{5plus1}.
This is the hallmark of fracton kinematics.
These constraints can be organized as higher-moment (``multipole'') global symmetries, as emphasized in continuum QFT treatments of fractons and related systems.
Our covariant formulation provides a compact way to encode the same conservation laws while keeping Poincar\'e covariance manifest. We do not attempt a systematic classification of the corresponding generalized symmetries here, but we will occasionally use this language when interpreting sources and conservation laws.

\section{Wilsonian structure: operator classification and basis relations}
\label{wilson}

The basic gauge-invariant building block is the generalized field strength $F_{\mu\nu\rho}$ \eqref{defF}, with $[F]=3$.
Local scalar densities are built from $F$ and derivatives, with indices contracted by $\eta_{\mu\nu}$ and, when needed, by $\epsilon^{\mu\nu\rho\sigma\lambda\tau}$.
We classify local gauge-invariant operators up to mass dimension eight, \emph{i.e.}\ including the leading four-derivative corrections to the two-derivative theory, and summarize the basis relations induced by total-derivative equivalences, Bianchi identities, and equations of motion.  For earlier general analyses of gauge procedures involving tensor gauge fields of rank higher than one, see also Ref.~\mbox{\cite{Singleton:2004je}}\hskip0pt.

At mass dimension six, locality and the strict power counting $\dim(\mathcal L)\le 6$ allow no gauge-invariant operators beyond the two quadratic structures already displayed in \eqref{Sgen}, namely $F_{\mu\nu\rho}F^{\mu\nu\rho}$ and $K_\mu K^\mu$ with $K_\mu\equiv -F_{\alpha\mu}{}^{\alpha}$.
This two-parameter quadratic sector is a key difference from ordinary Maxwell theory, where the two-derivative invariant is essentially unique, up to a $\theta$-term.
Accordingly, the leading nontrivial local deformations in the parity-even sector arise at mass dimension eight.
Concerning the parity-odd sector, the only odd local scalar density that can be built solely from the minimal gauge-invariant field strength is, up to an overall coefficient,
\be
\mathcal L_{\rm odd}^{(6)}=\epsilon^{\mu\nu\rho\sigma\lambda\tau}F_{\mu\nu\rho}F_{\sigma\lambda\tau},
\label{LepsFF}
\ee
which is a total derivative. As such, it is the higher-rank generalization of the ordinary topological $\theta$-term.
Consequently, the parity-odd sector does not furnish a nontrivial bulk deformation within the minimal field content.
\medskip
A convenient set of dimension-eight parity-even derivative operators is
\bea
\mathcal O_8^{(1)} &=& (\partial_\sigma F_{\mu\nu\rho})(\partial^\sigma F^{\mu\nu\rho}),\label{O81}\\
\mathcal O_8^{(2)} &=& (\partial_\mu F^{\mu\nu\rho})(\partial^\sigma F_{\sigma\nu\rho}),\label{O82}\\
\mathcal O_8^{(3)} &=& (\partial_\mu F_{\nu\rho\sigma})(\partial^\nu F^{\mu\rho\sigma}).\label{O83}
\eea
They are not all independent.
Up to total derivatives, which do not affect the action, the Bianchi identity \eqref{bianchi} implies that $\mathcal O_8^{(3)}$ can be reduced to a linear combination of $\mathcal O_8^{(1)}$ and $\mathcal O_8^{(2)}$.
We provide a compact derivation and a set of useful reduction identities in App.~\ref{O8reductions}.
Moreover, we will not make further use here of equations-of-motion reductions for $\mathcal O_8^{(2)}$; for the present discussion it is sufficient to keep the off-shell basis \eqref{O81}--\eqref{O83} and the total-derivative reduction stated below.
In a minimal derivative basis one may therefore retain $\mathcal O_8^{(1)}$ as the leading higher-derivative correction.
The removal of operators proportional to the equations of motion is standard in EFT, but in the present context it is important to verify that this basis choice is compatible with the mobility constraints once the theory is coupled to generic external sources.
We therefore keep the external source coupling
\begin{equation}
S_{\rm ext}=-\int d^6x\, A_{\mu\nu}J^{\mu\nu}\,,
\label{Sext}\end{equation}
which is gauge invariant provided the source satisfies
\begin{equation}
\partial_\mu\partial_\nu J^{\mu\nu}=0\,.
\label{ddJ}\end{equation}
A local field redefinition used to eliminate an EOM-squared operator in the bulk generates higher-derivative contact terms involving $J^{\mu\nu}$, but it does not alter the gauge-invariance condition above and hence does not modify the associated charge/dipole constraints discussed in Sect.~\ref{5plus1}.

A natural question is whether the restriction that $F_{\mu\nu\rho}$ derives from a symmetric potential,
\be
F_{\mu\nu\rho}=\partial_\mu A_{\nu\rho}+\partial_\nu A_{\mu\rho}-2\partial_\rho A_{\mu\nu},
\label{FfromA}
\ee
imposes additional \emph{algebraic} identities beyond symmetry in $(\mu,\nu)$ and the cyclic identity \eqref{cyclic}, potentially reducing the dimension of the quartic invariant space.
The answer is negative for purely algebraic invariants (such as $F^4$): the potential origin \eqref{FfromA} does not impose further local identities beyond symmetry in $(\mu,\nu)$ and the cyclic identity \eqref{cyclic}.
To see this, note that for \emph{constant} $A_{\mu\nu}$, the tensor $F_{\mu\nu\rho}$ as defined in \eqref{FfromA} vanishes identically; therefore any algebraic identity that would follow solely from the potential structure would require $F_{\mu\nu\rho}=0$ on a class of configurations where it need not vanish, which is not the case for a generic nonconstant field. Any identity that holds for all symmetric-potential tensors beyond symmetry in $(\mu,\nu)$ and the cyclic identity would thus already be encoded in those two properties, and no additional algebraic restriction on invariants built from $F$ without derivatives follows.
Extra reductions can only arise once derivatives are included, where the Bianchi identity, total-derivative equivalences, and equation-of-motion relations become effective.

\section{\texorpdfstring{$5+1$}{5+1} decomposition: generalized Maxwell equations and mobility constraints}\label{5plus1}

Our six-dimensional covariant formulation admits a transparent $5+1$ split in terms of gauge-invariant electric and magnetic variables. 
This makes the vacuum equations of motion \eqref{eom} and the source constraints \eqref{ddJ} particularly explicit, and provides a direct covariant route to charge and dipole conservation.

We now perform the split $\mu=(0,i)$ with $i=1,\dots,5$, under which the gauge transformation \eqref{intro_fract} becomes
\bea
\delta A_{00}&=&\partial_0^2\Lambda,\qquad
\delta A_{0i}=\partial_0\partial_i\Lambda,\qquad
\delta A_{ij}=\partial_i\partial_j\Lambda.
\eea
The Maxwell-like structure with nonvanishing magnetic sector is most transparent on the fractonic branch
\be
A_{\mu 0}=\partial_\mu\phi.
\label{fractonic_branch_main}
\ee
On this branch one has
\be
D_i\equiv \partial_0 A_{i0}-\partial_i A_{00}=0,
\qquad
E_{[ij]}=-\frac12\bigl(\partial_i A_{0j}-\partial_j A_{0i}\bigr)=0,
\label{branch_conditions_main}
\ee
so the electric tensor defined below is automatically symmetric.

In complete analogy with ordinary electrodynamics, we define the generalized electric tensor as the time--space component of the field strength,
\be
E_{ij}\equiv F_{0ij}=\partial_0 A_{ij}+\partial_i A_{0j}-2\partial_j A_{0i}.
\label{Edef}
\ee
Restricting to the branch \eqref{fractonic_branch_main} gives
\be
E_{ij}=\partial_0 A_{ij}-\partial_i\partial_j\phi=E_{ji}.
\label{Esymmetric_branch_main}
\ee
The magnetic sector is described by the rank-four tensor
\be
B_{klmn}\equiv \frac{1}{3}\,\epsilon_{klmij}\,F_{n}{}^{ij}
=\frac{1}{2}\,\epsilon_{klmij}\left(\partial^i A^j{}_{n}-\partial^j A^i{}_{n}\right)
=\epsilon_{klmij}\,\partial^i A^j{}_{n}.
\label{Bdef}
\ee
It obeys
\be
B_{klmn}=B_{[klm]n},
\label{Bsymm}
\ee
and the inverse relation
\be
\partial^i A^j{}_{n}-\partial^j A^i{}_{n}=\frac{1}{3!}\,\epsilon^{ijklm}B_{klmn}
\label{Binv}
\ee
shows that $B_{klmn}$ captures the full spatial curl sector in five dimensions.

Restricting the covariant field equations \eqref{eom} to the branch \eqref{fractonic_branch_main} yields the vacuum generalized Maxwell system. Choosing $(\alpha,\beta)=(0,j)$ gives the electric Gauss law
\be
\partial_i E^{ij}=0,
\label{Gauss}
\ee
while choosing $(\alpha,\beta)=(i,j)$ yields the Amp\`ere law
\be
\partial_0 E^{ij}+\frac{1}{2\cdot 3!}\left(\epsilon^{iklmn}\partial_k B_{lmn}{}^{j}+\epsilon^{jklmn}\partial_k B_{lmn}{}^{i}\right)=0.
\label{Ampere}
\ee
From the differential identity associated with the spatial curl, one obtains in turn the Faraday law
\be
\partial_0 B_{klmn}=\epsilon_{klmij}\,\partial^i E^{j}{}_{n},
\label{Faraday}
\ee
and the magnetic Gauss law
\be
\partial^k B_{klmn}=0.
\label{MagGauss}
\ee
Eqs.~\eqref{Gauss}--\eqref{MagGauss} form the complete generalized Maxwell system on the fractonic branch.

\subsection{Constraints and propagating modes}
\label{dof}

The $5+1$ split also clarifies the constrained nature of the theory. For recent general discussions of systematic methods for counting degrees of freedom, see e.g.\ Ref.~\cite{Hell:2026blj}.
{For the Maxwell-like representative \eqref{Sfract} ($\kappa=0$), the constrained structure can be stated in a fully explicit way: $A_{00}$ and $A_{0j}$ are fixed by the branch condition \eqref{fractonic_branch_main}, and variation with respect to $A_{0j}$ in the unreduced covariant action imposes the Gauss-type constraint \eqref{Gauss_constraint}. This is the starting point for the degree-of-freedom count given below.}
Using \eqref{Bdef} and its inverse \eqref{Binv}, one then finds the branch identity
\be
F_{\mu\nu\rho}F^{\mu\nu\rho}\Big|_{A_{\mu0}=\partial_\mu\phi}
=-6\,E_{ij}E^{ij}+B_{klmn}B^{klmn}.
\label{F2split_branch_main}
\ee
Hence the branch Lagrangian density is
\be
\mathcal L_{\rm branch}=\frac12\,E_{ij}E^{ij}-\frac1{12}\,B_{klmn}B^{klmn}.
\label{L_EB_main}
\ee
Equivalently, after the conventional rescaling
\be
\widehat B_{klmn}\equiv \frac{1}{\sqrt6}\,B_{klmn},
\label{Bhatdef_main}
\ee
one may write
\be
\mathcal L_{\rm branch}=\frac12\Big(E_{ij}E^{ij}-\widehat B_{klmn}\widehat B^{klmn}\Big).
\label{L_EBhat_main}
\ee
The canonical momentum conjugate to $A_{ij}$ is therefore
\be
\Pi^{ij} \equiv \frac{\partial\mathcal L_{\rm branch}}{\partial(\partial_0 A_{ij})}=E^{ij},
\ee
and the Gauss-type constraint takes the Hamiltonian form
\be
\mathcal G^j(x)\equiv \partial_i \Pi^{ij}(x)\approx 0.
\label{Gauss_constraint}
\ee
All physical statements in this section are gauge independent: on the branch the generalized fields $E_{ij}$ and $B_{klmn}$ are gauge invariant under \eqref{intro_fract}, and the generalized Maxwell system is written entirely in terms of these invariant quantities.  It is important, however, to distinguish this set of local relations from the actual gauge redundancy of the model. The Lagrangian gauge symmetry is scalar, $\delta A_{ij}=\partial_i\partial_j\Lambda$. We therefore do not interpret the family $\mathcal G^j\approx 0$ as providing five independent gauge generators in one-to-one correspondence with gauge parameters. Rather, only the longitudinally smeared combination of $\mathcal G^j$ is associated with the scalar gauge parameter, while the transverse combinations constrain the physical sector without enlarging the gauge orbit. Appendix~\ref{hamiltonian} now also makes this point fully explicit by decomposing a generic smearing into longitudinal and transverse parts.   For a comparison with the canonical constraint structure of related rank-two theories with scalar gauge symmetry, see~\mbox{\cite{Afxonidis:2024tph}}\hskip0pt; the difference in physical content reflects the different kinetic structures, as discussed in Sect.~\ref{setup}.

The vacuum equations imply relativistic dispersion. Combining the Amp\`ere law \eqref{Ampere} with the Faraday relation \eqref{Faraday} yields the wave equation
\be
\partial_0^2 E_{ij}-\Delta E_{ij}=0,
\label{wave_Eij}
\ee
together with the constraint $\partial_i E^{ij}=0$.

Plane waves for the generalized electric field may be written as $E_{ij}(t,\vec x)=e_{ij}\,e^{-i\omega t+i\vec k\cdot\vec x}$, with $\omega^2=\vec k^{\,2}$ and $k_i e^{ij}=0$.

\medskip
\noindent
Many continuum treatments of scalar-charge fracton phases in the condensed-matter literature adopt a nonrelativistic scaling in which the magnetic sector carries extra spatial derivatives, yielding a dynamical exponent \(z=2\) and \(\omega\sim k^2\).
Here we instead adopt, by construction, a two-derivative Lorentz-covariant action in \(5+1\) dimensions, so the propagating gauge modes obey the relativistic dispersion \(\omega^2=\vec k^{\,2}\).
This choice does \emph{not} affect the fractonic mobility restrictions, which are kinematical and follow from gauge invariance together with the source constraint \(\partial_\mu\partial_\nu J^{\mu\nu}=0\) derived below.

In momentum space, it is also useful to record the action of the residual scalar gauge symmetry on the spatial potential polarization $a_{ij}$:
\be
\delta a_{ij}=k_i k_j\,\Lambda.
\label{gauge_pol}
\ee
This condition removes only the scalar longitudinal piece of \(A_{ij}\), namely the component of the form \(\partial_i\partial_j\phi\).
It does not constrain the trace \(A^i{}_i\), so the isotropic trace component proportional to \(\delta_{ij}\) (the ``trace mode'') remains.

Moreover, the Gauss law \eqref{Gauss} implies, in momentum space,
\be
k_i E^{ij}=0 \quad\Rightarrow\quad k_i \varepsilon^{ij}=0,
\label{transverse}
\ee
which enforces transversality of the physical polarizations. Choosing coordinates such that $\vec k$ points along the $5$th axis, $k_i=(0,0,0,0,k)$, Eq.~\eqref{transverse} gives
\be
\varepsilon^{5j}=0 \qquad (j=1,\dots,5),
\label{eps5j}
\ee
so all components with one index along $\vec k$ vanish.
The remaining independent components are the symmetric tensor $\varepsilon^{ab}$ with $a,b=1,\dots,4$, i.e.\ a rank-2 symmetric tensor in the four-dimensional transverse space.

A symmetric $4\times 4$ tensor has $10$ independent components.
Summarizing, for nonzero momentum modes the   five relations $k_i e^{ij}=0$ enforce transversality of the gauge-invariant electric polarization, while the potential polarization is identified only under the scalar shift $a_{ij}\sim a_{ij}+k_i k_j\Lambda$. After imposing these transversality conditions and quotienting by this single scalar gauge redundancy, one is left with a symmetric transverse tensor in the four-dimensional little-group space . The remaining polarizations transform under the little group $SO(4)$ and decompose into trace and traceless parts,
\be
\varepsilon^{ab} = \left(\varepsilon^{ab}-\frac{1}{4}\delta^{ab}\varepsilon^c{}_c\right)
+\frac{1}{4}\delta^{ab}\varepsilon^c{}_c,
\ee
corresponding to a $9\oplus 1$ decomposition under $SO(4)$.

\medskip
\noindent
It is instructive to compare the propagating content of our rank-two gauge field with that of a massless graviton in six dimensions.
The traceless sector carries \(9\) polarizations, matching the physical degrees of freedom of a \(6\)D massless spin-\(2\) field.
In addition, our theory contains one extra scalar polarization, namely the trace \(A^i{}_i\), which cannot be removed by the scalar gauge symmetry \eqref{gauge_pol}.
This reflects the fact that \(\delta A_{\mu\nu}=\partial_\mu\partial_\nu\Lambda\) is only the longitudinal subset of the linearized diffeomorphisms \eqref{diff}. Accordingly,   unlike in ordinary linearized gravity,  the trace mode  is not removed and remains part of the physical spectrum .
In this work we instead focus on the Maxwell-like representative \eqref{Sfract}, where the symmetry remains scalar and the extra scalar mode is part of the physical spectrum.
A complementary Hamiltonian constraint analysis confirming the same propagating content and positivity on the physical subspace is presented in App.~\ref{hamiltonian}.
Altogether, the theory propagates ten local degrees of freedom in six dimensions.

\medskip
\subsection{Sources and mobility constraints}

Coupling the theory to a symmetric external source $J_{\mu\nu}$ via \eqref{Sext} modifies the field equations to
\begin{equation}
\partial_\mu F^{\alpha\beta\mu}=-J^{\alpha\beta} \, ,
\label{4.17}
\end{equation}
while gauge invariance under \eqref{intro_fract} imposes the covariant constraint
\begin{equation}
\partial_\mu\partial_\nu J^{\mu\nu}=0 \, .
\label{4.18}
\end{equation}
In $5+1$ form, defining $\rho\equiv J^{00}$, Eq.~\eqref{Jsplit} reads
\be
\partial_0^2 \rho + 2\partial_0\partial_i J^{0i} + \partial_i\partial_j J^{ij}=0.
\label{Jsplit}
\ee
Integrating over $\IR^5$ and assuming standard falloff conditions yields conservation of total charge
\be
Q=\int d^5x\,\rho,\qquad \frac{dQ}{dt}=0,
\ee
and, multiplying by $x^k$ and integrating by parts, conservation of dipole moment
\be
P^k=\int d^5x\,x^k\rho,\qquad \frac{dP^k}{dt}=0.
\ee
  In the covariant setting, the source constraint $\partial_\mu\partial_\nu J^{\mu\nu}=0$ has a further consequence that has no direct analogue in the non-relativistic case. Multiplying by $x^\nu$ and integrating, one finds that conservation holds not only for the spatial components $P^k$ but also constrains the temporal sector. Specifically, integrating by parts and using standard falloff conditions, one obtains that $\int d^5x\,J^{0k}$ is independently conserved:
\be
\frac{d}{dt}\int d^5x\,J^{0k}=0.
\ee
Together with the conservation of $Q$ and $P^k$, this implies that an isolated charged excitation is constrained to be immobile in the full spacetime sense: it cannot propagate as a worldline but is instead localized in both space and time. In this sense it is more accurately described as instanton-like rather than as a particle with restricted mobility. Sect.~\ref{conc} records this as the appropriate covariant restatement of the fractonic mobility constraint.

  A useful way to read these conservation laws is the following. If an isolated charge $q$ were translated by a displacement $a^k$, its contribution to the dipole moment would shift by $\Delta P^k=q\,a^k$. Unless $q=0$, even a rigid translation therefore changes the conserved quantity $P^k$, so isolated charged excitations cannot move as single particles. By contrast, a neutral dipolar bound state can translate without changing the total charge and with a fixed total dipole moment, because the opposite charges contribute compensating shifts.

 Therefore isolated charges cannot move without changing \(P^k\), while neutral composites can.   This immobility of isolated charges---together with the possibility of motion only for charge-neutral bound states---is the defining kinematical hallmark of fractonic behavior. For visual orientation, Fig.~\ref{fig:mobility_constraints} summarizes this contrast between an isolated charge, whose rigid translation would shift the conserved dipole moment, and a neutral dipole, whose rigid translation leaves it unchanged.

\begin{figure}[t]
\centering
\begin{tikzpicture}[x=1cm,y=1cm,line width=0.5pt,>=stealth,every node/.style={font=\small}]
  \begin{scope}[shift={(-4.2,0)}]
    \node[font=\normalsize] at (0.75,2.0) {(a) isolated charge};
    \fill (0,0.65) circle (2.3pt);
    \node[above left] at (0,0.65) {$q$};
    \draw[dashed,->] (0.18,0.78) -- (1.55,1.38);
    \draw (1.55,1.38) circle (2.3pt);
    \node at (0.86,1.64) {$a^k$};
    \node[align=center,text width=4.1cm] at (0.78,-0.15)
      {$\Delta P^k=q\,a^k\neq0$\\forbidden by dipole conservation};
  \end{scope}

  \begin{scope}[shift={(4.2,0)}]
    \node[font=\normalsize] at (0.35,2.0) {(b) neutral dipole};
    \draw (-1.25,0.65) circle (2.1pt);
    \draw (0.05,0.65) circle (2.1pt);
    \draw (-1.25,0.65) -- (0.05,0.65);
    \draw[dashed,->] (-0.2,-0.02) -- (1.2,-0.02);
    \node at (0.48,0.25) {$a^k$};
    \fill (0.55,0.65) circle (2.1pt);
    \fill (1.85,0.65) circle (2.1pt);
    \draw (0.55,0.65) -- (1.85,0.65);
    \node[above] at (0.55,0.65) {$+q$};
    \node[above] at (1.85,0.65) {$-q$};
    \node[align=center,text width=5.8cm] at (0.3,-0.78)
      {$\Delta Q=0,\qquad \Delta P^k=a^k\sum_n q_n=0$\\rigid translation preserves $Q$ and $P^k$};
  \end{scope}
\end{tikzpicture}
\caption{Schematic summary of the mobility constraints implied by simultaneous conservation of total charge and total dipole moment in the scalar-charge phase. Left: translating an isolated charge by $a^k$ shifts the dipole moment by $\Delta P^k=q\,a^k\neq0$. Right: a rigid translation of a neutral dipole leaves $\Delta Q=0$ and $\Delta P^k=a^k\sum_n q_n=0$, so the composite can move without violating the conservation laws.}
\label{fig:mobility_constraints}
\end{figure}

\medskip
\noindent
As a simple illustration, consider a localized point-charge trajectory $\rho(t,\vec x)=q\,\delta^{(5)}\bigl(\vec x-\vec x_0(t)\bigr)$, for which $Q=q$ and $P^k=q\,x_0^k(t)$. Dipole conservation then implies $\dot x_0^k(t)=0$ for $q\neq 0$, i.e.\ an isolated charge is immobile, while a neutral dipole with total $Q=0$ can move consistently with fixed $P^k$.  In the covariant language, this means that an isolated charged excitation is not described by a freely propagating worldline degree of freedom; rather, it is localized in spacetime in an instanton-like manner. We use this terminology only as an interpretation of the relativistic mobility constraint derived above.
  This makes precise, within the covariant field-theoretic setting, the standard schematic intuition that a single fractonic charge is pinned whereas a neutral dipole may propagate.

On the fractonic branch \eqref{fractonic_branch_main}, the sourced Maxwell system becomes
\bea
\partial_i E^{ij} &=& J^{0j},\label{4.20}\\
\partial_0 E^{ij}+\frac{1}{2\cdot 3!}\left(\epsilon^{iklmn}\partial_k B_{lmn}{}^{j}+\epsilon^{jklmn}\partial_k B_{lmn}{}^{i}\right) &=& -\,\tfrac12 J^{ij},\\
\partial_0 B_{klmn} &=& \epsilon_{klmij}\partial^i E^{j}{}_{n},\\
\partial^k B_{klmn} &=& 0.\label{4.23}
\eea

\section{Stress--energy tensor in \texorpdfstring{$d=6$}{d=6}: trace, improvement and the off-shell question}

\label{stress}

We define the stress--energy tensor by coupling the theory to a background metric and varying the action:
\be
T_{\mu\nu}(x)=-\frac{2}{\sqrt{-g}}\frac{\delta S}{\delta g^{\mu\nu}(x)}\Big|_{g=\eta}.
\label{Tdef}
\ee
 {This background-metric variation is used here only as a flat-space device to define the symmetric stress tensor and analyze its trace properties in the relativistic Maxwell-like theory. We do not assume, or need, a fully general curved-background completion preserving the scalar gauge symmetry on arbitrary geometries; for detailed discussions of the subtleties of coupling fracton gauge theories to curved backgrounds, see~\mbox{\cite{Slagle:2018lkl,Bidussi:2021nmp,Jain:2021ibh,Afxonidis:2025xge}}}. 
 For the Maxwell-like action \eqref{Sfract}, the physically relevant information for the present discussion is captured by its restriction to the fractonic branch \eqref{fractonic_branch_main}, for which \eqref{branch_conditions_main} holds.

\medskip
\noindent
Using the definitions \eqref{Edef}, \eqref{Bdef} and \eqref{branch_conditions_main}, one finds the branch decomposition
\be
F_{\mu\nu\rho}F^{\mu\nu\rho}\Big|_{A_{\mu0}=\partial_\mu\phi}
=-6\,E_{ij}E^{ij}+B_{klmn}B^{klmn}.
\label{F2EB}
\ee
Equivalently,
\be
\mathcal L_{\rm branch}=\frac12\,E_{ij}E^{ij}-\frac1{12}\,B_{klmn}B^{klmn}
=\frac12\Big(E_{ij}E^{ij}-\widehat B_{klmn}\widehat B^{klmn}\Big),
\label{L_EB_main_stress}
\ee
where \(\widehat B_{klmn}\equiv B_{klmn}/\sqrt6\). Thus the generalized Maxwell structure with symmetric electric field and nonzero magnetic field is realized precisely on the fractonic branch.

At the level of the energy density, one finds, up to the standard superpotential ambiguity,
\be
T_{00}\Big|_{A_{\mu0}=\partial_\mu\phi}=\frac12\,E_{ij}E^{ij}+\frac1{12}\,B_{klmn}B^{klmn}
+\partial^\rho\partial^\sigma Y_{0\rho0\sigma}
=\frac12\Big(E_{ij}E^{ij}+\widehat B_{klmn}\widehat B^{klmn}\Big)+\partial^\rho\partial^\sigma Y_{0\rho0\sigma}.
\label{T00EB}
\ee
In particular, positivity is manifest on the branch. A complementary canonical analysis in the same sector is given in App.~\ref{hamiltonian}.

\medskip
\noindent
Using the Maxwell-like action together with the equations of motion, one obtains the following trace identity, which depends only on the Maxwell-like $F^2$ structure, up to a total derivative:
\be
T^\mu{}_{\mu}=-\frac{d-6}{12}\,F_{\alpha\beta\gamma}F^{\alpha\beta\gamma}+\partial_\mu V^\mu\ ,
\label{trace_general}
\ee
where $V^\mu$ can be written in terms of the superpotential $Y_{\mu\rho\nu\sigma}$ as
\be
V^\mu\equiv \partial_\sigma Y^{\rho\mu}_{\ \ \rho}\ .
\label{defV}\ee
Hence in $d=6$,
\be
T^\mu{}_{\mu}=\partial_\mu V^\mu.
\label{trace6}
\ee
Notice that the disappearance of the $F^2$ term at $d=6$ is fixed by mass dimensions and does not rely on a gauge choice.

\medskip
\noindent
The distinction between scale invariance and conformal invariance is well known. In a local QFT, scale invariance allows the trace to be a total derivative,
\(T^\mu{}_{\mu}=\partial_\mu V^\mu\), where \(V^\mu\) is often referred to as a virial current.
Conformal invariance is more restrictive: it requires that this virial current be removable by a local improvement, i.e.\ that \(V^\mu\) can be written as \(V^\mu=(d-1)\partial^\mu\Phi\) for some local scalar \(\Phi\), so that an improved stress--energy tensor can be made traceless off shell; see e.g.\ Refs.~\cite{Polchinski:1987dy,Jack:1990eb,Fortin:2011ks,Dymarsky:2014zja,Nakayama:2013is} for general discussions.

In the present theory, Eq.~\eqref{trace6} realizes the scale-invariant pattern in $d=6$ under standard boundary conditions.
However, \eqref{trace6} is obtained using the field equations, whereas the improvement problem asks a sharper, genuinely off-shell question: can the trace be removed by a \emph{local identity} at the level of the stress--energy tensor?
Concretely, we consider local CCJ improvements \cite{Callan:1970ze} of the form \eqref{impr_ansatz}, with a local scalar operator \(\Phi\).
Unless stated otherwise, we further restrict to \(\Phi\) that are manifestly gauge invariant and built within the minimal field content, so that any obstruction is an off-shell statement intrinsic to the theory rather than an artifact of gauge-variant representatives.

The standard local improvement in flat space takes the form
\be
T'_{\mu\nu}=T_{\mu\nu}+(\partial_\mu\partial_\nu-\eta_{\mu\nu}\Box)\Phi,
\label{impr_ansatz}
\ee
with some local scalar $\Phi$, under which
\be
T'{}^\mu{}_{\mu} = T^\mu{}_{\mu}-(d-1)\Box\Phi.
\label{trace_shift}
\ee
In $d=6$, off-shell tracelessness would require
\be
\partial_\mu V^\mu = 5\,\Box\Phi
\label{need_phi}
\ee
as a local identity for some local $\Phi$, where the coefficient $5=d-1$ follows from \eqref{trace_shift} evaluated at $d=6$.

In $d=6$ one has $[F]=[K]=3$.
To keep locality and manifest gauge invariance, $\Phi$ must be a local scalar polynomial in $F_{\mu\nu\rho}$, $K_\mu$, and derivatives.
Power counting then severely restricts the possibilities:
any scalar containing at least two factors of $F$ and/or $K$ has mass dimension $\ge 6$, while a dimension-$4$ scalar can only be \emph{linear} in the gauge-invariant data and must carry derivatives.
Up to total derivatives, the only Lorentz scalar of dimension $4$ is
\be
\Phi_{\rm inv}\;\propto\; \partial_\mu K^\mu.
\label{Phi_gi_candidate}
\ee
However, $\Phi_{\rm inv}$ is itself a total derivative and is linear in the fields. By contrast, as shown explicitly in \eqref{Vdiv_app}, a representative virial divergence $\partial_\mu V^\mu$ is bilinear in the fields, up to terms proportional to the equations of motion. Therefore, the identity \eqref{need_phi} cannot be satisfied off shell within the minimal local polynomial, manifestly gauge-invariant operator algebra generated by $F_{\mu\nu\rho}$, $K_\mu$, and derivatives.
Equivalently, removing a quadratic virial term by a CCJ improvement would require a dimension-4 scalar $\Phi$ that is quadratic in the basic field, but the only such local candidate is $A_{\mu\nu}A^{\mu\nu}$, which is gauge variant. The obstruction is therefore an off-shell statement about the nonexistence of a local manifestly gauge-invariant improvement within the minimal field content, and should not be conflated with quantum anomalies. A compact derivation of a convenient explicit choice of $V^\mu$, together with a detailed check that the gauge-invariant candidate \eqref{Phi_gi_candidate} cannot solve \eqref{need_phi}, is given in App.~\ref{virial} (see in particular \eqref{Vchoice_app}, \eqref{Vdiv_app}, and \eqref{BoxPhi_app}).

Although the manifestly gauge-invariant off-shell improvement is obstructed, in $d=6$ we still have \eqref{trace6}, and one can construct an on-shell traceless representative by allowing $\Phi$ to be gauge variant or by using EOM-proportional identities.
A simple explicit choice is
\be
\Phi \equiv \frac{1}{4}A_{\mu\nu}A^{\mu\nu},
\label{Phi_choice}
\ee
which is the simplest local scalar of dimension $4$ built from $A_{\mu\nu}$; it is not gauge invariant, but this is admissible for an on-shell representative.
This yields
\be
T'{}^\mu{}_{\mu}
=\partial_\mu \widetilde V^\mu
+\frac{1}{2}\,A_{\alpha\beta}\,\partial_\mu F^{\alpha\beta\mu},
\label{trace_onshell}
\ee
where
\be
\widetilde V^\mu \equiv V^\mu - 5\,\partial^\mu\Phi
= -\,A_{\nu\rho}F^{\mu\nu\rho} - \frac{5}{4}\,\partial^\mu\!\left(A_{\nu\rho}A^{\nu\rho}\right)
\label{Vtildedef}
\ee
collects the total-derivative contributions from the original virial current \eqref{Vchoice_app} and from the CCJ improvement term.
Thus $T'{}^\mu{}_{\mu}$ vanishes on shell (up to a total derivative), making precise the sense in which the $d=6$ reference theory is classically scale invariant in flat space under standard boundary conditions.

\section{Conclusions and outlook}
\label{conc}

In this work we developed a six-dimensional covariant formulation of scalar-charge fracton electrodynamics based on a symmetric rank-2 tensor gauge field \(A_{\mu\nu}(x)\) with scalar gauge symmetry
\[
\delta A_{\mu\nu}=\partial_\mu\partial_\nu \Lambda \, .
\]
Our main motivation was to identify a relativistic setting in which the characteristic kinematics of fracton systems can be expressed in a local and manifestly covariant language, while at the same time keeping under control the stress--energy tensor and the organization of gauge-invariant deformations. In this sense, \(d=6\) is singled out as the natural reference dimension for the Maxwell-like theory: the two-derivative kinetic term is marginal, the gauge parameter can be taken dimensionless, and the resulting model provides a natural scale-invariant starting point for this tensor gauge structure. The role of six spacetime dimensions in our discussion is not meant to be directly phenomenological. Rather, \(d=6\) plays the role of a natural covariant reference point, in which locality, gauge symmetry and scaling are simultaneously most transparent, while lower-dimensional realizations are more naturally interpreted as effective descriptions away from the critical dimension.

Although the present six-dimensional formulation is primarily motivated by its structural
simplicity and by the marginality of the two-derivative kinetic term, it is worth noting
several broader contexts in which this gauge structure may naturally appear.

First, rank-two symmetric tensor fields play a central role in string theory, where the
massless excitations of the closed string include both the graviton and additional mixed-symmetry
tensor fields.  
Even though the present scalar-charge gauge symmetry does not coincide with
linearized diffeomorphisms, it fits naturally within the general taxonomy of higher-rank gauge
fields that arise in effective descriptions of compactified or constrained string backgrounds.

Second, in effective field theories with spontaneously broken higher-form symmetries or in
non-Lorentzian limits relevant for elasticity dualities, symmetric tensor fields frequently
emerge as low-energy degrees of freedom.  
In such contexts the conservation of multipole moments
may appear either as an emergent constraint or as a remnant of microscopic subsystem symmetries.

Finally, tensor gauge structures similar to the one analyzed here arise in continuum limits of
lattice models with restricted mobility, generalized elasticity theories, and higher-moment
hydrodynamics.  
Although the six-dimensional model developed in this work is not intended as a
direct phenomenological theory, its covariance and locality make it a useful reference point for
constructing effective theories in lower dimensions or for interpreting higher-moment
generalized global symmetries within a relativistic field-theoretic framework.

A central physical outcome of the analysis is that the characteristic mobility constraints of scalar-charge fracton theories arise here directly from gauge invariance. Coupling the theory to a symmetric external source \(J_{\mu\nu}(x)\) enforces the covariant constraint
\[
\partial_\mu\partial_\nu J^{\mu\nu}=0 \, ,
\]
and in the \(5+1\) decomposition this becomes the conservation of total charge and dipole moment. As a consequence, isolated charges are immobile, whereas neutral dipolar composites can propagate. In this way, the defining fractonic behavior is not imposed as an additional nonrelativistic rule, but follows from the covariant gauge principle itself. {In the relativistic setting, the additional conservation of $\int d^5x\,J^{0k}$ means that an isolated charged excitation is localized in spacetime rather than carried by a propagating worldline; this is the precise sense in which the covariant mobility constraint is more appropriately described as instanton-like.} The same \(5+1\) split also reorganizes the equations of motion into a generalized Maxwell system for gauge-invariant electric and magnetic variables, making the physical content of the theory especially transparent.

A second result concerns the stress--energy tensor. For the Maxwell-like representative, its trace takes the universal form
\[
T^\mu{}_\mu = -\frac{d-6}{12} F_{\alpha\beta\gamma}F^{\alpha\beta\gamma} + \partial_\mu V^\mu \, .
\]
Accordingly, in \(d=6\) the explicit \(F^2\) contribution disappears and the trace reduces to a total derivative, implying classical scale invariance in flat space under standard boundary conditions. At the same time, our analysis shows that this statement is more rigid than in ordinary Maxwell theory. Within the minimal local gauge-invariant operator algebra, the available dimension-four scalar candidates are not sufficient to remove the virial term by a local, manifestly gauge-invariant improvement. Thus, although an on-shell traceless representative can still be constructed, the off-shell problem exhibits a genuine obstruction within the minimal field content. This points to a nontrivial interplay between gauge symmetry, locality and scale invariance which appears to be intrinsic to the six-dimensional theory rather than an artifact of a particular stress--tensor representative.

From a Wilsonian viewpoint, the six-dimensional model also provides a clean organizing center for covariant fracton-like gauge dynamics. The marginal sector is highly constrained, while the leading irrelevant deformations can be reduced to a small set of independent structures once Bianchi identities, integrations by parts and equations of motion are taken into account. Although we have not addressed here the quantum \(\beta\)-function or possible interacting UV completions, the present construction offers a concrete local framework in which such questions can be posed systematically.

Several directions   merit  further investigation. A first question is whether the obstruction to a local manifestly gauge-invariant traceless representative persists in BRST-based formulations or in enlarged field contents. A second issue concerns boundary conditions and defects, where the \(5+1\) formulation may help clarify the realization of generalized symmetries and the associated physical observables. Related boundary-induced structures in covariant rank-two theories were analyzed in~\cite{Bertolini:2023sqa}. More broadly, it would be interesting to understand whether the six-dimensional model discussed here should be viewed simply as a distinguished Gaussian reference point, or rather as the starting point for a wider relativistic formulation of fracton quantum field theory.  In this broader perspective, the present model adds a symmetric-tensor fracton example to the wider six-dimensional landscape in which scale and conformal structures often acquire a particularly rigid form, even though the field content and symmetry principles differ substantially from those of six-dimensional SCFTs.

\appendix

\section{Useful identities for the generalized magnetic tensor}
\label{dictionary}

For convenience we collect here the identities most often used in the main text. On the fractonic branch \eqref{fractonic_branch_main}, the electric variable is the symmetric tensor
\be
E_{ij}\equiv F_{0ij}=\partial_0 A_{ij}+\partial_i A_{0j}-2\partial_j A_{0i}
=\partial_0 A_{ij}-\partial_i\partial_j\phi,
\label{Esym_lit_app}
\ee
and the magnetic variable is the rank-four tensor
\be
B_{klmn}\equiv \frac{1}{3}\,\epsilon_{klmij}\,F_{n}{}^{ij}=\epsilon_{klmij}\,\partial^i A^j{}_{n}.
\label{Bsym_lit_app}
\ee
By construction,
\be
B_{klmn}=B_{[klm]n},
\ee
and the inverse relation reads
\be
\partial^i A^j{}_{n}-\partial^j A^i{}_{n}=\frac{1}{3!}\,\epsilon^{ijklm}B_{klmn}.
\label{FinvB_app}
\ee
A useful quadratic identity is
\be
\bigl(\partial_i A_j{}^{n}-\partial_j A_i{}^{n}\bigr)\bigl(\partial^i A^{j}{}_{n}-\partial^j A^{i}{}_{n}\bigr)=\frac{1}{3!}\,B_{klmn}B^{klmn},
\label{B2id_app}
\ee
while the branch conditions imply
\be
D_i=\partial_0 A_{i0}-\partial_i A_{00}=0,\qquad E_{[ij]}=0.
\label{branch_conditions_app}
\ee
Hence
\be
F_{\mu\nu\rho}F^{\mu\nu\rho}\Big|_{A_{\mu0}=\partial_\mu\phi}=-6\,E_{ij}E^{ij}+B_{klmn}B^{klmn},
\ee
and therefore Eqs.~\eqref{F2EB} and \eqref{T00EB} follow directly.

\section{Hamiltonian constraint analysis in \texorpdfstring{$5+1$}{5+1}}
\label{hamiltonian}

This appendix provides a streamlined constraint analysis supporting Sect.~\ref{dof} on the fractonic branch \eqref{fractonic_branch_main}.
We work in flat space with \(\eta_{\mu\nu}=\mathrm{diag}(-,+,+,+,+,+)\) and use the Maxwell-like action \eqref{Sfract}.

On the branch \(A_{\mu0}=\partial_\mu\phi\), the gauge-invariant combinations
\be
D_i\equiv \partial_0 A_{i0}-\partial_i A_{00},\qquad E_{[ij]}=-\frac12\bigl(\partial_i A_{0j}-\partial_j A_{0i}\bigr)
\ee
vanish identically, so the quadratic action reduces to
\be
\mathcal L_{\rm branch}=\frac12\,E_{ij}E^{ij}-\frac1{12}\,B_{klmn}B^{klmn}
=\frac12\Big(E_{ij}E^{ij}-\widehat B_{klmn}\widehat B^{klmn}\Big),
\label{L_EB_app}
\ee
with \(\widehat B_{klmn}\equiv B_{klmn}/\sqrt6\).

Because $A_{ij}$ is symmetric, the momentum conjugate to it is
\be
\Pi^{ij}\equiv \frac{\partial\mathcal L_{\rm branch}}{\partial(\partial_0 A_{ij})}
=E^{ij}.
\label{PiDef}
\ee
Eliminating the velocities in favor of \(\Pi^{ij}\) yields the canonical Hamiltonian density
\be
\mathcal H_{\rm branch}=\frac12\,\Pi_{ij}\Pi^{ij}+\frac1{12}\,B_{klmn}B^{klmn}-A_{0j}\,\partial_i\Pi^{ij}.
\label{Hcanon}
\ee
Equivalently,
\be
\mathcal H_{\rm branch}=\frac12\Big(\Pi_{ij}\Pi^{ij}+\widehat B_{klmn}\widehat B^{klmn}\Big)-A_{0j}\,\partial_i\Pi^{ij}.
\ee
The Euler--Lagrange equation obtained by varying the unreduced covariant action with respect to \(A_{0j}\) reproduces the Gauss law \eqref{Gauss}; after restricting to the branch this becomes \(\partial_i E^{ij}=0\) in vacuum. Therefore the Hamiltonian density is manifestly nonnegative on the physical subspace, exactly as stated in the main text.

\section{Dimension-eight operator reductions}
\label{O8reductions}

This appendix collects the reduction identities used in Sect.~\ref{wilson} for the dimension-eight quadratic operators \eqref{O81}--\eqref{O83}.
We work modulo total derivatives and use the corrected Bianchi identity \eqref{bianchi}.

\medskip
\noindent
\textbf{Reduction of $\mathcal O_8^{(3)}$.}
Starting from
\be
\mathcal O_8^{(3)}\equiv (\partial_\mu F_{\nu\rho\sigma})(\partial^\nu F^{\mu\rho\sigma}),
\ee
it is convenient to introduce the antisymmetric combination
\be
X_{\mu\nu\rho}\equiv F_{\mu\rho\nu}-F_{\nu\rho\mu},
\qquad X_{\mu\nu\rho}=-X_{\nu\mu\rho},
\label{Xdef_app}
\ee
which obeys the ordinary Bianchi identity
\be
\partial_\sigma X_{\mu\nu\rho}+\partial_\mu X_{\nu\sigma\rho}+\partial_\nu X_{\sigma\mu\rho}=0.
\label{Xbianchi_app}
\ee
Using \eqref{Xdef_app} together with the symmetry of $F$ in its first two indices and the cyclic identity \eqref{cyclic}, one may rewrite the derivative basis in terms of $X$ and obtain, after integrations by parts,
\be
\mathcal O_8^{(3)}=\mathcal O_8^{(1)}-\mathcal O_8^{(2)}+\partial_\mu J_{(8)}^{\mu},
\label{O83_reduction}
\ee
for a local current $J_{(8)}^{\mu}$. In particular, modulo total derivatives,
\be
\mathcal O_8^{(3)}\equiv \mathcal O_8^{(1)}-\mathcal O_8^{(2)}.
\ee
This is the only off-shell reduction needed in the main text.

\medskip
\noindent
\textbf{On-shell corollary.}
For the purposes of the main text, the off-shell reduction \eqref{O83_reduction} is the only relation we need. In particular, no additional on-shell reduction is assumed here beyond the Maxwell equation \eqref{eom}.

\section{Virial current and the gauge-invariant improvement check}
\label{virial}

This appendix collects explicit formulae underlying the trace discussion in Sect.~\ref{stress}.
{We make the off-shell status of the improvement problem fully explicit by separating the bilinear virial-current identity from the linear gauge-invariant candidate $\Phi_{\rm inv}\propto\partial_\mu K^\mu$.}
Our goal is not to fix a unique representative, since superpotentials allow many equivalent choices, but rather to provide
(i) a convenient explicit virial current $V^\mu$ whose divergence has the same off-shell schematic structure as the $d=6$ trace up to terms proportional to the equations of motion, and
(ii) a direct check that the manifestly gauge-invariant candidate $\Phi_{\rm inv}\propto\partial_\mu K^\mu$ cannot solve the off-shell improvement condition \eqref{need_phi}.

\medskip
\noindent
\textbf{A useful identity.}
Expanding the field strength definition \eqref{defF} one finds the algebraic identity
\bea
F_{\mu\nu\rho}F^{\mu\nu\rho}
&=&\bigl(\partial_\mu A_{\nu\rho}+\partial_\nu A_{\mu\rho}-2\partial_\rho A_{\mu\nu}\bigr)
\bigl(\partial^\mu A^{\nu\rho}+\partial^\nu A^{\mu\rho}-2\partial^\rho A^{\mu\nu}\bigr)\nn\\
&=&6\Big[(\partial_\mu A_{\nu\rho})(\partial^\mu A^{\nu\rho})-(\partial_\mu A_{\nu\rho})(\partial^\nu A^{\mu\rho})\Big].
\label{F2_expand}
\eea
From this it follows immediately that
\be
(\partial_\mu A_{\nu\rho})F^{\mu\nu\rho}=\frac{1}{6}\,F_{\mu\nu\rho}F^{\mu\nu\rho}.
\label{dAF_halfF2}
\ee
Consequently,
\bea
\partial_\mu\bigl(A_{\nu\rho}F^{\mu\nu\rho}\bigr)
&=&(\partial_\mu A_{\nu\rho})F^{\mu\nu\rho}+A_{\nu\rho}\,\partial_\mu F^{\mu\nu\rho}\nn\\
&=&\frac{1}{6}\,F_{\mu\nu\rho}F^{\mu\nu\rho}+A_{\nu\rho}\,\partial_\mu F^{\mu\nu\rho}.
\label{AF_div}
\eea

In $d=6$ the trace takes the form \eqref{trace6} after using the equations of motion to remove terms proportional to $\partial_\mu F^{\mu\nu\rho}$. 
Equation~\eqref{AF_div} shows that the bilinear current
\be
V^\mu\equiv -\,A_{\nu\rho}F^{\mu\nu\rho}
\label{Vchoice_app}
\ee
has divergence
\be
\partial_\mu V^\mu=-\frac{1}{6}\,F_{\mu\nu\rho}F^{\mu\nu\rho}-A_{\nu\rho}\,\partial_\mu F^{\mu\nu\rho}.
\label{Vdiv_app}
\ee
Thus, up to terms proportional to the equations of motion, $\partial_\mu V^\mu$ is bilinear in the fields.
Different stress--tensor representatives correspond to shifting $V^\mu$ by identically conserved currents and/or shifting $T_{\mu\nu}$ by superpotentials; none of these operations changes the key fact that $\partial_\mu V^\mu$ is bilinear in the basic field for the free theory.

In $d=6$ the improvement condition \eqref{need_phi} requires $[\Phi]=4$.
Since $[F]=[K]=3$, any manifestly gauge-invariant local polynomial scalar built from $F$, $K$, and derivatives must be either:
(a) at least two factors of $F$ and/or $K$ (hence dimension $\ge 6$), or
(b) exactly one factor of $F$ or $K$ accompanied by derivatives.
But a Lorentz scalar linear in $F_{\mu\nu\rho}$ does not exist, since its indices cannot be saturated without introducing additional tensor factors, so the only possibility is linear in $K_\mu$.
Up to total derivatives, within the minimal class of local manifestly gauge-invariant polynomials, the only scalar of dimension $4$ is therefore $\partial_\mu K^\mu$, namely Eq.~\eqref{Phi_gi_candidate}.

For any choice $\Phi=c\,\partial_\mu K^\mu$, one has
\be
\Box\Phi=c\,\Box\partial_\mu K^\mu,
\label{BoxPhi_app}
\ee
which is linear in the fields, since $K_\mu$ is linear in $A_{\mu\nu}$.
By contrast, as shown explicitly by \eqref{Vdiv_app}, a representative divergence $\partial_\mu V^\mu$ is bilinear in the fields, modulo equations of motion.
Therefore the equality $\partial_\mu V^\mu=5\Box\Phi$ cannot hold as an off-shell local operator identity within the minimal local manifestly gauge-invariant CCJ-type ansatz.
The obstruction established here is therefore an off-shell statement about the nonexistence of a local manifestly gauge-invariant improvement within the minimal field content, and does not conflict with the existence of an on-shell traceless representative.

\section*{Acknowledgments}

I thank Erica Bertolini and Alberto Blasi for instructive discussions.


\end{document}